\documentclass[twocolumn]{aastex631}
\usepackage[varg]{txfonts}
\usepackage{graphicx} 
\usepackage{url} 

\begin{document}
\title{Protostellar Disk Formation Regimes: Angular Momentum Conservation versus Magnetic Braking}

\author[0000-0003-1412-893X]{Hsi-Wei Yen}
\affiliation{Academia Sinica Institute of Astronomy \& Astrophysics,
11F of Astronomy-Mathematics Building, AS/NTU, No.1, Sec. 4, Roosevelt Rd, Taipei 10617, Taiwan}

\author[0000-0003-3497-2329]{Yueh-Ning Lee}
\affiliation{Department of Earth Sciences, National Taiwan Normal University, Taipei 116059, Taiwan}
\affiliation{Center of Astronomy and Gravitation, National Taiwan Normal University, Taipei 116059, Taiwan}
\affiliation{Physics Division, National Center for Theoretical Sciences, Taipei 106319, Taiwan}

\correspondingauthor{Hsi-Wei Yen}
\email{hwyen@asiaa.sinica.edu.tw}

\shortauthors{Yen \& Lee}

\begin{abstract}
Protostellar disks around young protostars exhibit diverse properties, with their radii ranging from less than ten to several hundred au. 
To investigate the mechanisms shaping this disk radius distribution, 
we compiled a sample of 27 Class 0 and I single protostars with resolved disks and dynamically determined protostellar masses from the literature. 
Additionally, we derived the radial profile of the rotational to gravitational energy ratio in dense cores from the observed specific angular momentum profiles in the literature. 
Using these observed protostellar masses and rotational energy profile, we computed theoretical disk radii from the hydrodynamic and non-ideal magnetohydrodynamic (MHD) models in \citet{LeeYN21,LeeYN24} and generated synthetic samples to compare with the observations. 
In our theoretical model, the disk radii are determined by hydrodynamics when the central protostar+disk mass is low. 
After the protostars and disks grow and exceed certain masses, the disk radii become regulated by magnetic braking and non-ideal MHD effects. 
The synthetic disk radius distribution from this model matches well with the observations. 
This result suggests that hydrodynamics and non-ideal MHD can be dominant in different mass regimes (or evolutionary stages) depending on the rotational energy and protostar+disk mass.  
This model naturally explains the rarity of large ($>$100 au) disks and the presence of very small ($<$10 au) disks. 
It also predicts that the majority of protostellar disks have radii of a few tens of au, as observed.
\end{abstract}


\section{Introduction}
Protostellar disks are often observed around young protostars and are potential sites of planet formation \citep{Manara18, Hsieh24}. 
Observations reveal that most protostellar disks have radii of a few tens of au, while large ($>$100 au) disks are rare \citep{Tsukamoto23}. 
Besides, several very small ($<$10 au) disks also present \citep{Tobin20a}. 
This distribution is different from the expectation from hydrodynamics, where large disks can easily form \citep{Hennebelle16, LeeYN21}. 
These results are often interpreted as magnetic fields suppressing disk growth via magnetic braking \citep{Yen15, Maury19, Tobin20a},
although the presence of small disks may also be explained from pure hydrodynamics if the rotational energy in dense cores is low \citep{Harsono14}. 

In the ideal magnetohydrodynamics (MHD) limit, magnetic braking is highly efficient at suppressing disk formation and growth, 
and no disks larger than 10 au form in ideal MHD simulations \citep{Allen03, Mellon08}.
This contradicts the observations. 
Several mechanisms have been proposed to alleviate magnetic braking and enable the formation of sizable protostellar disks with radii of a few tens of au. 
These mechanisms include very weak magnetic fields \citep{Mellon08}, misaligned magnetic fields with the rotational axis \citep{Joos12, Li13}, turbulence \citep{Santos12, Seifried12}, dissipation of protostellar envelopes \citep{Machida11}, and non-ideal MHD effects \citep{Inutsuka10, Masson16, Zhao16}.

Observationally, sizable disks are often deeply embedded in dense cores \citep{Tobin20a}, 
and their radii are independent on the turbulent velocity and magnetic field strengths and orientations in their environments \citep{Yen21b, Yen24}. 
Thus, these are less likely the primary mechanisms for the formation of sizable disks in magnetized dense cores,  
and non-ideal MHD effects remain a key candidate mechanism in disk formation and growth \citep{Wurster20, Tsukamoto23}.

Several theoretical studies on non-ideal MHD effects on disk radii have predicted the relationship between disk radii and the masses of central protostar+disk systems \citep{Hennebelle16, Tsukamoto20, LeeYN21, LeeYN24}. 
With high-resolution and sensitivity observations, 
disk rotation has been resolved in molecular-line emission in a sample of young protostars, enabling dynamical determination of protostellar masses \citep{Tobin12}. 
These results allow for a statistical comparison between observations and theories of the relationship between disk radii and the masses of central protostar+disk systems, to investigate non-ideal MHD effects on disk formation and growth.

In this paper, we compile a sample of 27 Class 0 and I single protostars with resolved disks and dynamically determined protostellar masses from the literature. 
We then apply those theoretical predictions, compute theoretical disk radii, and generate synthetic samples to compare with the observed disk radii. Through this comparison between the theoretical and observed distributions of protostellar disk radii, 
we discuss the relative importance of hydrodynamics and non-ideal MHD in the disk formation and growth process.

\section{Theoretical disk radius}\label{sec:theory}
The radius of a protostellar disk embedded in an infalling protostellar envelope can be estimated from the equilibrium radius at the boundary between the disk and envelope in a quasi-stationary state, as discussed in detail by \citet{Hennebelle16} and \citet{LeeYN21,LeeYN24}. 
This equilibrium radius is determined by the net flux of angular momentum transported by mass infall toward the disk and transferred outward by magnetic braking. 
In this paper, we adopt the models in \citet{LeeYN21,LeeYN24} and follow their formulation to compute the theoretical disk radii. 
The key concepts and formulas presented in \citet{LeeYN21,LeeYN24} are summarized below.

\subsection{Hydrodynamics}
From hydrodynamics, the angular momentum of infalling material in a protostellar source is conserved, and there is no magnetic braking. 
In the hydrodynamic model in \citet{LeeYN21}, the theoretical disk radius ($R_{\rm hydro}$) increases with the amount of mass that has been accreted onto the central protostar+disk system ($M$) and the ratio of rotational to gravitational energy ($\beta$) that infalling material possesses. 
From Equation~14 in \citet{LeeYN21}, $R_{\rm hydro}$ can be expressed as, 
\begin{equation}\label{eq:Rhydro}
R_{\rm hydro} = 111\ {\rm au} \left[\frac{\beta}{0.02}\right]^{\frac{1}{2}} \left[\frac{M}{0.1\ M_\sun}\right]. 
\end{equation}

\subsection{Non-ideal MHD}
From MHD, the angular momentum of infalling material is partially transferred outward via magnetic braking, primarily due to the tension of azimuthal magnetic fields. 
The strength of the azimuthal magnetic fields is determined by induction and diffusion, regulated by non-ideal MHD effects. 
Thus, the equilibrium radius can be evaluated from the balance between mass infall and magnetic braking, and the balance between magnetic induction and diffusion. 
Ambipolar diffusion is expected to be the dominant non-ideal MHD effect in protostellar envelopes, considering their typical density \citep{Masson16, Zhao16}.
In addition, this equilibrium radius does not depend on the initial rotational energy in the dense core. 
Fast rotation leads to stronger magnetic braking, as the two can self regulate each other \citep{Hennebelle16}.
Therefore, the theoretical disk radius from non-ideal MHD increases with the resistivity of ambipolar diffusion and the mass of the central protostar+disk system, and decreases with the magnetic field strength \citep{Hennebelle16, LeeYN21,LeeYN24}.

The resistivity of ambipolar diffusion depends on the density, ionization fraction and magnetic field strength in protostellar envelopes. 
Given a chemical model, the ionization fraction can be parameterized as a function of density \citep{Marchand16}.
Additionally, the magnetic field strength is influenced by the resistivity of ambipolar diffusion and density. 
Therefore, in the non-ideal MHD model in \citet{LeeYN24}, the resistivity of ambipolar diffusion and magnetic field strength are parameterized together as a function of density, or equivalently, the enclosed mass.

Consequently, in the non-ideal MHD model in \citet{LeeYN24}, the theoretical disk radius ($R_{\rm MHD}$) with its magnetic field inclined from the disk normal axis by $i$ degrees is derived to be a function of mass of the central protostar+disk system. 
From Equations C12 and C13 in \citet{LeeYN24}, $R_{\rm MHD}$ can be expressed as, 
\begin{eqnarray}\label{eq:incli}
\left({R_{\rm MHD} \over r_{\rm v}}\right)^{1-{9\over4}\alpha} \cos{i} + \left({R_{\rm MHD} \over r_{\rm h}}\right)^{{3\over 2}-{9\over4}\alpha} \sin{i}  = 1,
\end{eqnarray}
where
\begin{eqnarray}\label{eq:MHD}
 \left\{ 
 \begin{array}{l} 
r_{\rm v} =  4.50\ {\rm au} \left[{M \over 0.1 M_\odot}\right]^{1\over 13} \\
r_{\rm h} =  334\ {\rm au} \left[{M \over 0.1 M_\odot}\right]^{21\over 33} \\
\alpha = 0.3
\end{array} \right.
~{\rm or}~ 
 \left\{ 
 \begin{array}{l} 
r_{\rm v} =  37.8\ {\rm au} \left[{M \over 0.1 M_\odot}\right]^{1\over 4} \\
r_{\rm h} =  199\ {\rm au} \left[{M \over 0.1 M_\odot}\right]^{1\over 2} \\
\alpha = 0
\end{array} \right..
\end{eqnarray}
Here, the dependence of $R_{\rm MHD}$ on the resistivity of ambipolar diffusion and magnetic field strength are absorbed into its dependence on the mass. 
$\alpha$ is the power-law index of the ion and H$_2$ density relation \citep{Marchand16}.
$r_{\rm v}$ and $r_{\rm h}$ represent the disk radii in two extreme magnetic field configurations, vertical ($i = 0\arcdeg$) or horizontal ($i = 90\arcdeg$) fields with respect to the disk plane. 
The choice of the set of $r_{\rm v}$, $r_{\rm h}$, and $\alpha$ for computing the theoretical disk radius depends on the corresponding density, 
and the two regimes are separated at a density of 10$^{9}$ cm$^{-3}$.
For a given inclination of the magnetic field, the set leading to a larger radius should be adopted for the calculation.

The analytical calculations of theoretical disk radii from non-ideal MHD models with similar approaches have been examined with numerical simulations,
and the difference between the theoretical disk radii from analytical calculations and non-ideal MHD simulations is within a factor of two \citep{Hennebelle16}.
Thus, we assume that the uncertainty in the theoretical disk radius from the non-ideal MHD model in \citet{LeeYN24} to be a factor two in our subsequent analysis.

As detailed in \citet{LeeYN21,LeeYN24}, Equation \ref{eq:Rhydro} and \ref{eq:MHD} include numerical factors on the order of unity. 
Nevertheless, the uncertainties in observational measurements of core rotation and mass of protostar+disk systems are often larger than this order of magnitude (Sec.~\ref{sec:obs}). 
Therefore, these numerical factors in the equations are negligible compared to the observational uncertainties. 
For simplicity, they are all set to one in this paper.

\section{Observational data}\label{sec:obs}
The theoretical disk radii depend on core rotation, the mass of central protostar+disk systems, and the orientations of global magnetic fields in the initial dense cores (Sec.~\ref{sec:theory}).
In this section, we compile observational measurements of these parameters to evaluate theoretical disk radii for comparisons with observations.

\subsection{Dense core rotation}
For a given mass of a central protostar+disk system, its theoretical disk radius from hydrodynamics is determined by the ratio of rotational to gravitational energy of the material accreted from its initial dense core (Eq.~\ref{eq:Rhydro}). 
This ratio $\beta$ is often assumed to be a constant of 0.02 \citep[e.g.,][]{Hennebelle16}. 
This is based on the observations of the global velocity gradients on 0.06--0.6 pc scales in a sample of dense cores and the assumptions that dense cores are rigid-body rotating and have uniform density \citep{Goodman93}.
Under these assumptions, $\beta$ remains constant throughout the dense cores, independent on their core mass and size. 
However, these assumptions are likely not valid. 

\begin{deluxetable*}{lccccc}
\tablecaption{Specific angular momentum profiles in dense cores} \label{tab:core}
\tablewidth{0pt}
\tablehead{
\multicolumn{6}{c}{$j(r) = j_0 \times (\frac{r}{0.1\ {\rm pc}})^p$ km s$^{-1}$ pc} \\
\hline
\colhead{Reference} & \colhead{$j_0$} & \colhead{$p$} & \colhead{Radial range (pc)} & \colhead{$N_{\rm sample}$} & \colhead{Tracer} 
}
\startdata
\citet{Goodman93} & $10^{-2.3\pm0.3}$ & 1.6$\pm$0.2 & 0.06--0.6 & 23 & NH$_3$ \\
\citet{Punanova18} & 10$^{-1.9\pm1.3}$ & 2.4$\pm$0.9 & 0.02--0.07 & 13 & N$_2$H$^+$ \\
\citet{Pineda19} & $10^{-1.2\pm0.2}$ & 1.8$\pm$0.04 & 0.005--0.06 & 3 & NH$_3$ \\
\citet{Pandhi23} & $10^{-2.4\pm0.5}$ & 1.8$\pm$0.1 & 0.006--0.12 & 329 & NH$_3$ \\
\hline
Average & $10^{-2.4\pm0.5}$ & 1.8$\pm$0.2
\enddata
\tablecomments{The radial range is the spatial scale probed to measure the specific angular momentum profile. $N_{\rm sample}$ is the number of the sample sources, and the tracer is the molecular line observed in the study. The averaged profile is computed by averaging the four observed profiles weighted by their sample sizes. The error propagation is included. The sample size in \citet{Pandhi23} is dominant over the others. Thus, the averaged profile is similar to that in \citet{Pandhi23}, and the uncertainty in the power-law index increases because of the diverse values from these profiles.}
\end{deluxetable*}

Recent observations have probed core rotation in a large sample of dense cores \citep[e.g.,][]{Caselli02,Tatematsu16,Gaudel20}.
Radial profiles of specific angular momentum in dense cores are inferred either by resolving radial velocity profiles in individual dense cores \citep{Pineda19} or by comparing global velocity gradients and core sizes of a sample of dense cores \citep{Goodman93, Punanova18, Pandhi23}.

\begin{figure}[h]
\includegraphics[width=0.46\textwidth]{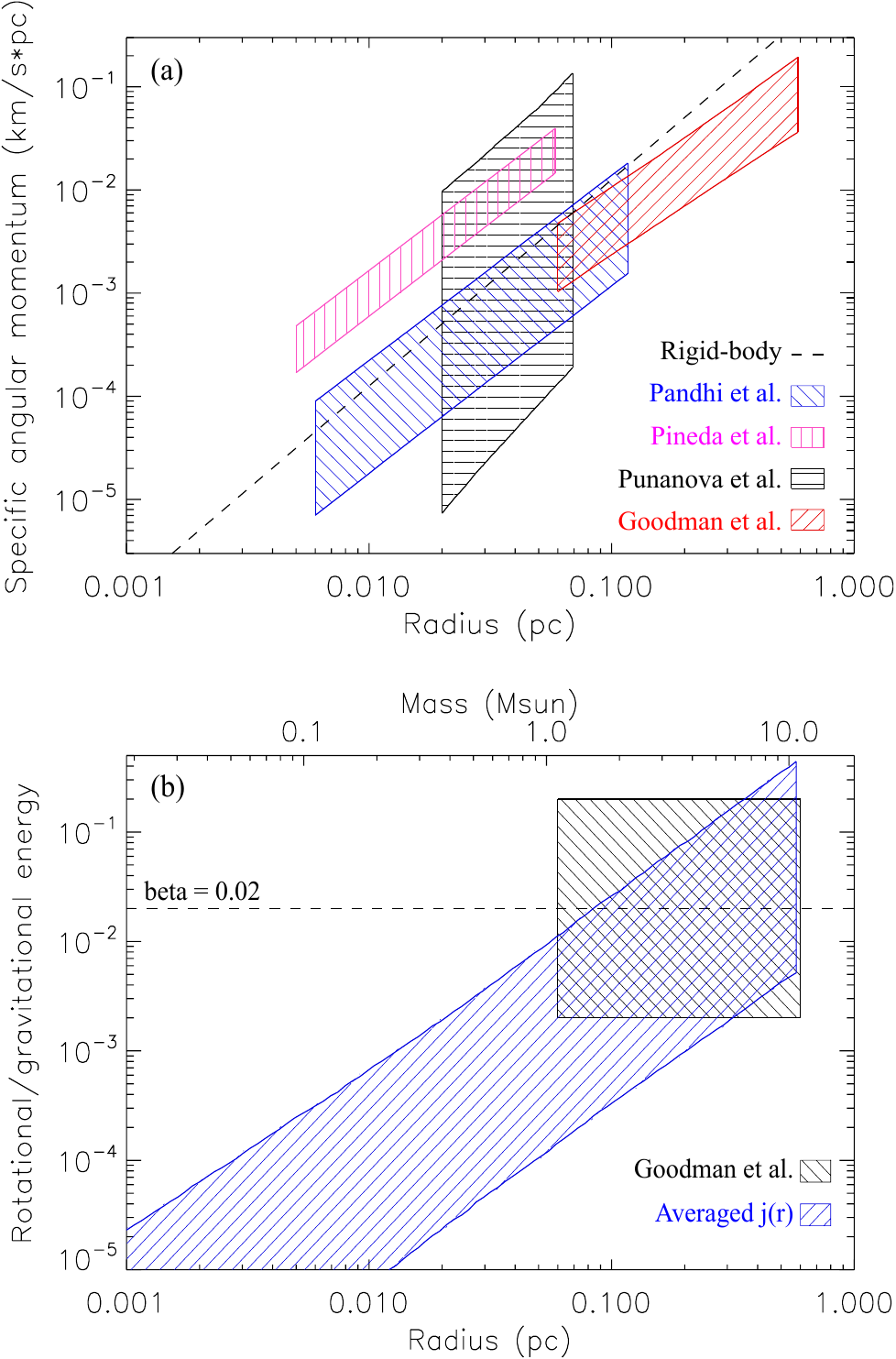}
\caption{(a) Radial profiles of specific angular momentum in dense cores inferred from observations. Shade areas in different patterns and colors present the results from different literatures as labeled in the panel. The horizontal and vertical extents of the shaded areas denote the spatial ranges probed by those observations and the uncertainties in their measurements, respectively. A dashed line demonstrates the slope of the specific angular momentum profile of a dense core in rigid-body rotation. (b) Radial profiles of the ratio of rotational to gravitational energy in dense cores. Blue shaded area presents the profile derived from the averaged specific angular momentum profile of four observational results shown in panel (a), and its vertical extents present the uncertainty. Black shaded area presents the ranges of the sizes and ratios of rotational to gravitational energy in dense cores observed by \citet{Goodman93} for comparison. A horizontal dash line denotes a constant ratio of 0.02. The corresponding enclosed mass for a given radius is labeled at the top of the panel.}
\label{fig:jr}
\end{figure}

In Table~\ref{tab:core}, we compile a list of observationally inferred specific angular momentum profiles from the literature, presented in Fig.~\ref{fig:jr}.
These observations cover different spatial scales of dense cores, and  
the specific angular momenta among dense cores at a given spatial scale can differ by more than one order of magnitude.
The power-law indices of these specific angular momentum profiles range from 1.6 to 2.4. 
They are slightly deviated from the expectation of rigid-body rotation having a power-law index of 2.

We assume dense cores initially to be singular isothermal spheres (SIS) having a density ($\rho$) profile described by 
\begin{equation}
\rho = \frac{c_{\rm s}^2}{2\pi G} r^{-2},
\end{equation}
where $c_{\rm s}$ is the sound speed and $G$ is the gravitational constant \citep{Shu77}, 
because the density in dense cores is typically centrally concentrated and is not uniform \citep{Motte01}. 
Though it is still a simplified assumption. 
For our calculations, $c_{\rm s}$ is adopted to be 0.2 km s$^{-1}$, approximately the sound speed at 10 K.
Using this density profile and the average specific angular momentum profile from observations, weighted by sample sizes (Table 1), 
we compute the rotational and gravitational energy as a function of radius. 
Then, the profile of the ratio of rotational to gravitational energy in initial dense cores is derived to be
\begin{equation}\label{eq:beta}
\beta = 10^{-3.0\pm1.2} \times (\frac{M}{1\ M_\sun})^{1.6\pm0.5},  
\end{equation}
where the radius enclosing 1 $M_\sun$ is 0.055 pc. 
This equation can be expressed as a function of either radius or enclosed mass because the enclosed mass is linearly proportional to radius in a SIS. 
At radii of 0.06--0.6 pc, $\beta$ from Eq.~\ref{eq:beta} is similar to the range of $\beta$ observed in a sample of dense cores in \citet{Goodman93}, around 0.02, 
and it decreases rapidly as the radius decreases (Fig.~\ref{fig:jr}b).

\subsection{Protostellar disk sample}
\subsubsection{Sample selection}
To compare theoretical and observed disk radii, we compile a list of observational measurements of disk radii, protostellar masses, and disk masses of Class 0 and I protostars from the literature (Table \ref{tab:disk}).
Our sample consists of 27 single protostars, meaning that there is no companion within 2000 au. 
Their disks have been resolved in the millimeter continuum emission, and signs of Keplerian rotation have been detected in molecular-line emission, with interferometric observations.

We select protostars without companions within 2000 au to avoid the influence of tidal interaction on disk sizes \citep{Manara19}. 
A clear correlation between disk radii and separations to companions has been observed in the Orion star-forming region, and this correlation becomes insignificant when the separation is larger than 2000 au  \citep{Yen21b}.
We note that some selected protostars may be unresolved close binaries. 
In such cases, given their small separation, they can be treated as a point source in the dynamics on the disk scale, 
and thus it does not affect our calculations in Sec.~\ref{sec:theory}. 

\subsubsection{Disk radius}
The disk radii of our sample sources are all estimated by two-dimensional Gaussian fitting to the central compact components of the observed continuum emission for a uniform comparison.\footnote{Although disk rotation is observed in molecular lines in our sample sources, it remains challenging to resolve the transition from the protostellar envelopes to the disks and determine the disk radii from the gas kinematics in the majority of our sample sources in these studies.}
The full-width-half-maximum widths of the fitted deconvolved Gaussian functions, adopted from the literature, are listed in Table \ref{tab:disk}. 
The disk radius adopted to be the 2$\sigma$ width of the major axis of the fitted deconvolved Gaussian function. 
It can represent the radius enclosing 90\% of the total flux of the disk ($R_{90\%}$). 

We note that the disk radii measured in the continuum are often smaller than those measured in the molecular lines \citep{Ansdell18}.
In case of the Class 0 and I protostars, such as L1527~IRS and TMC-1A, their disk radii in the continuum are 30\%--70\% of those determined from gas kinematics \citep{Aso15, Aso17, Hoff23}. 
The analysis of synthetic images of protostellar disks, generated from theoretical simulations and radiative transfer calculations, demonstrates that 
the disk radii can be under or overestimated in continuum emission compared to the actual disk radii in the theoretical simulations \citep{Aso20, Tung24}.
The ratio between the continuum and actual disk radii depends on the angular resolution of observations and contamination from surrounding protostellar envelopes \citep{Tung24}. 
These imaging simulations suggest that the uncertainty in $R_{90\%}$ is approximately 0.3 dex compared to the actual disk radius.  
This discrepancy between the continuum and actual disk radii is also comparable to those found in L1527~IRS and TMC-1A.
Therefore, in our sample, we assume that the uncertainty in the disk radii is 0.3 dex. 
The uncertainty in the Gaussian fitting to the continuum emission is typically less than 10\% \citep[e.g.,][]{Ohashi23}, so it is negligible compared to the systematic uncertainty of 0.3 dex.

\begin{figure}[h]
\includegraphics[width=0.46\textwidth]{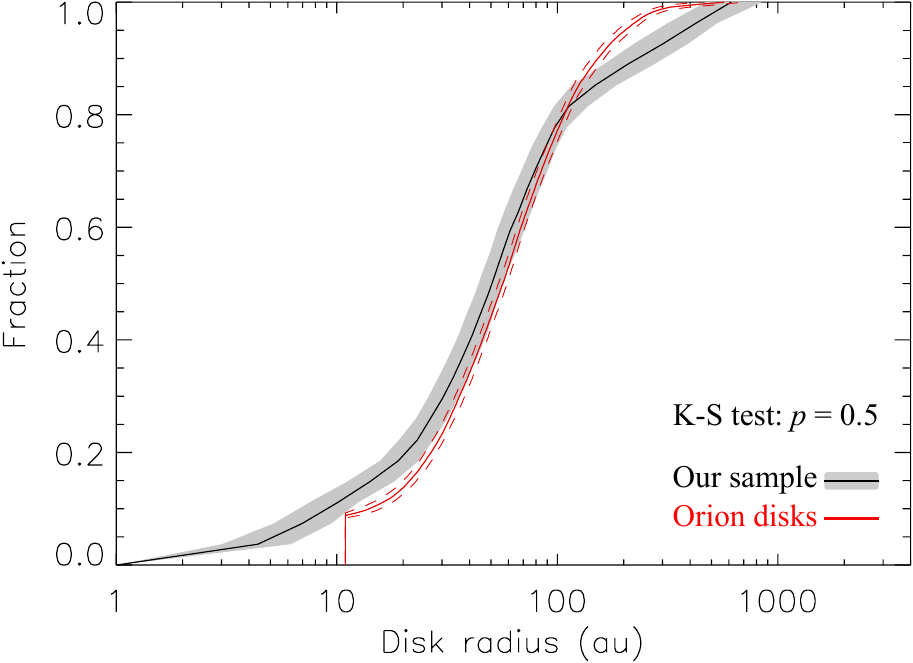}
\caption{Cumulative distributions of the disk radii in our sample (black) and in Orion measured with the ALMA observations \citep[red;][]{Tobin20a}. The gray shaded area red dashed lines present the uncertainties of the disk size distributions. In the Orion sample, only disks without companions within 2000 au are included, the same as our sample. 18 out of 227 Orion disks included here are unresolved, and the minimum observed disk radius is 11 au, so its cumulative distribution starts at 11 au with a fraction of 0.08. The K-S test of these two distributions results in a $p$ value of 0.5.}
\label{fig:Rd}
\end{figure}

Figure \ref{fig:Rd} presents the cumulative distribution of the disk radii in our sample in comparison with that in the Orion star-forming region measured with the ALMA observations at a 0\farcs1 resolution \citep{Tobin20a}. 
For consistency, we adopt the same definition of disk radius and only select disks without companions within 2000 au from the Orion data, 
and the same uncertainty of 0.3 dex is assumed.
Among the selected 227 Orion disks, 18 of them are unresolved, and the minimum observed radius is 11 au.
To derive the cumulative distributions, we carried out 1000 iterations, and varied the measurements within their uncertainties in each iteration. 
Then, the 16\%, 50\%, and 84\% percentiles were extracted from these 1000 iterations and presented in Fig.~\ref{fig:Rd}.
We performed a Kolmogorov-Smirnov (K-S) test on the cumulative distributions of the disks with radii larger than 11 au in our and Orion samples.
The resulting $p$ value is 0.5. 
Thus, the disk radius distribution in our sample is similar to that in the Orion star-forming region, suggesting no bias toward any specific disk radius.

\subsubsection{Protostellar and disk mass}
In Eq.~\ref{eq:Rhydro} and \ref{eq:MHD}, $M$ is the sum of the protostellar and disk masses.
The protostellar masses in our sample are estimated by fitting or comparing Keplerian rotation models to the observed velocity profiles extracted from position--velocity diagrams or velocity channel maps in the molecular-line emission. 
The values and their uncertainties, adopted from the literature, are listed in Table \ref{tab:disk}.
In principle, the mass derived from Keplerian velocity at a given radius is the enclosed mass, including both protostellar mass and part of disk mass. 
The enclosed disk mass increases with radius. 
The fraction of disk mass included in these protostellar mass estimates in the literature is not clear. 
This uncertainty is  accounted for through the uncertainty in the disk mass, as described below.

The disk masses are estimated from the fluxes of the (sub-)millimeter continuum emission in the disks, adopted from the literature and listed in Table \ref{tab:disk}.  
These quoted disk masses likely have uncertainties of a factor of two to three due to the assumed dust absorption coefficients and temperatures.
Nevertheless, the disk masses in 23 out of 27 sample sources are less than 10\% of their protostellar masses. 
Only BHR71~IRS1, BHR7, Lupus~3~MMS, and HH~211 have significant disk masses of 25\% 42\%, 44\%, and 96\% compared to their estimated protostellar masses, respectively.
Therefore, we do not expect that the uncertainty in the disk masses significantly affects our subsequent analysis because of the small ratios of disk to protostellar masses for the majority of the sample sources.  
For simplicity, we adopt an uncertainty of 100\% for all the disk masses. 

Then the protostellar and disk masses are added up to plug in Eq.~\ref{eq:Rhydro} and \ref{eq:MHD} in our analysis.
With the adopted uncertainty of 100\% for the disk masses, the lower limit corresponds to the case where the mass estimated from the Keplerian velocity includes the protostellar and entire disk masses. 

\begin{deluxetable*}{lccccccccccc}
\tablecaption{Sample sources and their properties} \label{tab:disk}
\tablewidth{0pt}
\tablehead{
\colhead{Source} & \colhead{RA} & \colhead{Dec}  & \colhead{Distance} & \colhead{Class} & \colhead{$T_{\rm bol}$} & \colhead{$L_{\rm bol}$} & \colhead{FWHM} & \colhead{$R_{\rm disk}$} & \colhead{$M_\star$} & \colhead{$M_{\rm disk}$} & \colhead{Reference} \\ 
\colhead{} & \colhead{(J2000)} & \colhead{(J2000)} & \colhead{(pc)} & \colhead{} & \colhead{(K)} & \colhead{($L_\sun$)} & \colhead{(mas)} & \colhead{(au)} & \colhead{($M_\sun$)} & \colhead{($M_\sun$)} & \colhead{}
}
\startdata
IRAS~15398$-$3359 & 15:43:02.23 & $-$34:09:07.0 & 155 & 0 &  50 &   1.4 &   29 &   4 & 0.06$\pm$0.04 & 1.7$\times$10$^{-3}$ & 1, 2\\
L1521F & 04:28:38.95 & +26:51:35.0 & 140 & 0/I &  20 &   0.1 &   58 &   7 & 0.18$\pm$0.05 & 4.0$\times$10$^{-4}$ & 3, 4, 5\\
BHR71~IRS2 & 12:01:34.01 & $-$65:08:48.1 & 176 & 0 &  39 &   1.1 &   49 &   7 & 0.26$\pm$0.01 & 3.9$\times$10$^{-3}$ & 1, 6\\
IRAS~16253$-$2429 & 16:28:21.62 & $-$24:36:24.3 & 139 & 0 &  42 &   0.2 &  107 &  13 & 0.15$^{+0.02}_{-0.03}$ & 2.1$\times$10$^{-3}$ & 1, 7 \\
IRAS~04166+2706  & 04:19:42.51 & +27:13:35.8 & 156 & 0 &  61 &   0.4 &  138 &  18 & 0.27$\pm$0.12 & 1.5$\times$10$^{-2}$ & 1, 8, 9 \\
L483 & 18:17:29.94 & $-$04:39:39.6 & 200 & 0 &  50 &  10.0 &  160 &  27 & 0.15$\pm$0.05 & 8.8$\times$10$^{-4}$ & 10, 11\\
IRAS~16544$-$1604 & 16:57:19.64 & $-$16:09:24.0 & 151 & 0 &  50 &   0.9 &  207 &  27 & 0.15$\pm$0.01 & 1.0$\times$10$^{-2}$ & 1, 12\\
IRAS~04169+2702 & 04:19:58.48 & +27:09:56.8 & 156 & I & 163 &   1.5 &  216 &  29 & 1.4$\pm$0.7 & 2.0$\times$10$^{-2}$ & 1, 8, 13\\
TMC-1A & 04:39:35.20 & +25:41:44.2 & 137 & I & 183 &   2.3 &  259 &  30 & 0.68$\pm$0.06 & 4.2$\times$10$^{-2}$ & 1, 14 \\
HH~211\tablenotemark{a} & 03:43:56.81 & +32:00:50.2 & 295 & 0 &  27 &   3.0 &  127 &  32 & 0.08 & 7.7$\times$10$^{-2}$ & 15, 16\\
Elias~29 & 16:27:09.44 & $-$24:37:19.3 & 137 & I & 391 &  13.6 &  312 &  36 & 0.9$\pm$0.1 & 1.7$\times$10$^{-3}$ & 17\\
L1448-C\tablenotemark{a} & 03:25:38.88 & +30:44:05.3 & 288 & 0 &  47 &  10.5 &  160 &  39 & 1.50$\pm$0.72 & 1.1$\times$10$^{-1}$ & 18, 19, 20\\
BHR71~IRS1 & 12:01:36.45 & $-$65:08:49.4 & 176 & 0 &  66 &  10.0 &  279 &  42 & 0.46$\pm$0.01 & 1.1$\times$10$^{-1}$ & 1, 6\\
R~CrA~IRS~5N & 19:01:48.48 & $-$36:57:15.4 & 147 & 0 &  59 &   1.4 &  374 &  47 & 0.29$\pm$0.11 & 1.9$\times$10$^{-2}$ & 1, 21\\
Oph~IRS~63 & 16:31:35.65 & $-$24:01:30.1 & 132 & I & 348 &   1.3 &  426 &  48 & 0.5$\pm$0.2 & 5.0$\times$10$^{-2}$ & 1, 22\\
HH~24~NE & 05:46:08.92 & $-$00:09:56.1 & 427 & I & 147 &  22.6 &  141 &  51 & 2.39$^{+0.23}_{-0.11}$ & 6.8$\times$10$^{-2}$ & 23, 24\\
$[$GY92$]$197 & 16:27:05.24 & $-$24:36:29.6 & 139 & I & 120 &   0.2 &  440 &  52 & 0.23$\pm$0.02 & 2.3$\times$10$^{-2}$ & 25 \\
L1527~IRS & 04:39:53.88 & +26:03:09.4 & 140 & 0 &  41 &   1.3 &  444 &  53 & 0.4$\pm$0.1 & 3.0$\times$10$^{-2}$ & 1, 26\\
Lupus~3~MMS & 16:09:18.09 & $-$39:04:53.3 & 162 & 0 &  39 &   0.3 &  410 &  56 & 0.15 & 6.6$\times$10$^{-2}$ & 27\\
GSS30~IRS3 & 16:26:21.72 & $-$24:22:51.1 & 138 & 0 &  50 &   1.7 &  550 &  64 & 0.35$\pm$0.09 & 2.1$\times$10$^{-2}$ & 1, 28\\
HH~212 & 05:43:51.41 & $-$01:02:53.2 & 413 & 0 &  53 &  14.0 &  200 &  70 & 0.27$\pm$0.05 & 1.4$\times$10$^{-3}$ & 29, 30\\
TMC1 & 04:41:12.70 & +25:46:34.8 & 140 & I & 101 &   0.9 &  800 &  95 & 0.54$^{+0.20}_{-0.10}$ & 3.9$\times$10$^{-3}$ & 31 \\
HOPS-370 & 05:35:27.63 & $-$05:09:34.4 & 392 & 0/I &  72 & 314.0 &  340 & 113 & 2.5$\pm$0.5 & 3.5$\times$10$^{-2}$ & 32\\
BHR7 & 08:14:23.33 & $-$34:31:03.7 & 400 & 0 &  51 &   9.3 &  520 & 177 & 1.0$\pm$0.4 & 4.2$\times$10$^{-1}$ & 33 \\
IRAS~04302+2247 & 04:33:16.50 & +22:53:20.2 & 160 & I &  88 &   0.4 & 2149 & 292 & 1.6$\pm$0.4 & 4.2$\times$10$^{-2}$ & 1, 34\\
Per-emb-8 & 03:44:44.98 & +32:01:35.2 & 295 & 0 &  43 &   2.5 & 1220 & 306 & 2.2$\pm$0.6 & 1.1$\times$10$^{-1}$ & 35 \\
L1489~IRS & 04:04:43.08 & +26:18:56.1 & 146 & I & 213 &   3.4 & 3913 & 485 & 1.7$\pm$0.2 & 9.3$\times$10$^{-3}$ & 36\\
\enddata
\tablenotetext{a}{$L_{\rm bol}$, $M_\star$, or $M_{\rm disk}$ from the literature are updated with more recent estimate of the distance to the source \citep{Tobin16,Dzib18,Zucker18,Tobin20a}.}
\tablecomments{$L_{\rm bol}$ and $T_{\rm bol}$ are bolometric luminosity and temperature. FWHM is the full-width-half-maximum of the major axis of the fitted deconvolved 2D Gaussian function to the continuum emission, and $R_{\rm disk}$ is the disk radius, defined as the 2$\sigma$ width of this Gaussian function. $M_\star$ and $M_{\rm disk}$ are protostellar and disk masses. }
\tablerefs{$^{1}$\citet{Ohashi23}; $^{2}$\citet{Thieme23}; $^{3}$\citet{Hsieh17}, $^{4}$\citet{Tokuda17}; $^{5}$\citet{Tokuda24}; $^{6}$\citet{Gavino24}; $^{7}$\citet{Aso23}; $^{8}$\citet{Yen24};  $^{9}$ Phuong et al.~(in prep.); $^{10}$\citet{Oya17}; $^{11}$\citet{Jacobsen19}; $^{12}$\citet{Kido23}; $^{13}$Han et al.~(in prep.); $^{14}$\citet{Aso15}; $^{15}$\citet{Lee18}; $^{16}$\citet{Yen23}; $^{17}$\citet{Oya19}; $^{18}$\citet{Hirano10}; $^{19}$\citet{Maury19}; $^{20}$\citet{Maret20}; $^{21}$\citet{Sharma23}; $^{22}$\citet{Flores23}; $^{23}$\citet{Tobin20a}; $^{24}$\citet{Reipurth23}; $^{25}$\citet{Villarmois19}; $^{26}$\citet{Hoff23}; $^{27}$\citet{Yen17}; $^{28}$\citet{Santamaria24}; $^{29}$\citet{Lee14}; $^{30}$\citet{Lee17}; $^{31}$\citet{Harsono14}; $^{32}$\citet{Tobin20b}; $^{33}$\citet{Tobin18}; $^{34}$\citet{Lin23}; $^{35}$\citet{Lin24}; $^{36}$\citet{Yamato23}
}
\end{deluxetable*}

\section{Analysis and results}\label{sec:results}
\begin{figure}[h]
\includegraphics[width=0.47\textwidth]{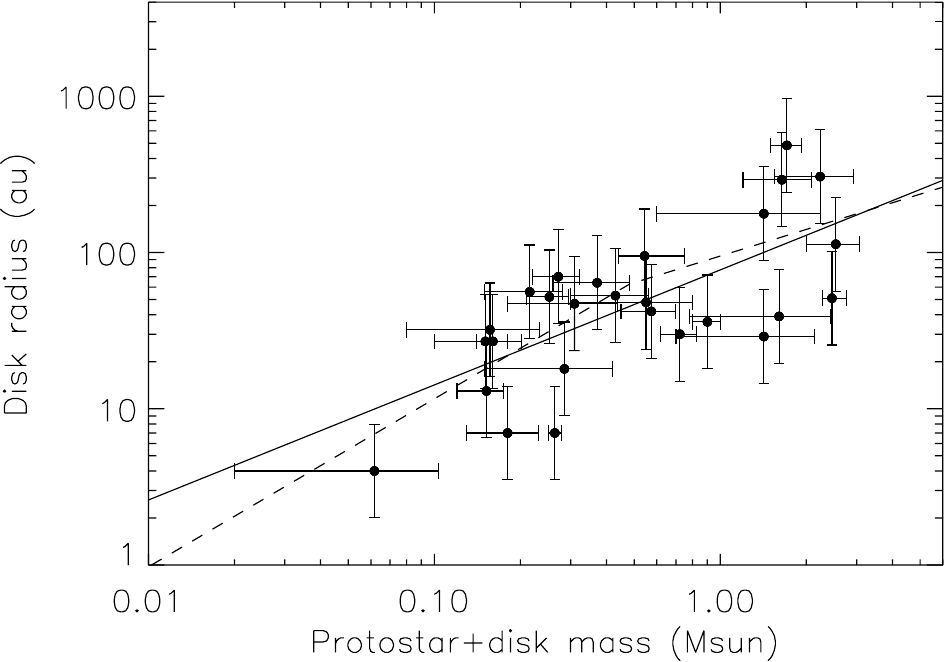}
\caption{Observed disk radius and protostar+disk mass distribution in our sample. The solid line presents the single power-law fitting to the data points, which results in a power-law index of 0.7$\pm$0.1. The dashed line presents the double power-law fitting, where the power-law indices are 1.1 and 0.6 at masses lower and higher than 0.5 $M_\sun$, respectively. Statistically, the fitted single power-law function better represents the data.}
\label{fig:ob_rd}
\end{figure}

Figure \ref{fig:ob_rd} presents the distribution of the observed disk radii and protostar+disk masses in our sample. 
The disk radii range from 4 to 485 au, and the protostar+disk masses range from 0.06 to 2.5 $M_\sun$.
The power-law fitting to these data points results in a power-law index of 0.7$\pm$0.1. 
If a double power-law function is adopted, the fitting results in power-law indices of 1.1 and 0.6 at masses lower and higher than 0.5 $M_\sun$, respectively. 
The reduced $\chi^2$ and Bayesian information criterion of the single power-law fitting are 2.6 and 71, 
while those of the double power-law fitting are 3.6 and 96, respectively. 
Thus, statistically the fitted single power-law function better represents the data.

The power-law index of the dependence of disk radius on protostar+disk mass from the hydrodynamic model in \citet{LeeYN21} is 1.8$\pm$0.25 (Eq.~\ref{eq:Rhydro} and \ref{eq:beta}), while that from the non-ideal MHD model in \citet{LeeYN24} ranges from 0.25 to 0.8, depending on magnetic field orientaitons (Eq.~\ref{eq:MHD}).  
The observed dependence based on the single power-law fitting is shallower than the prediction from the hydrodynamic model, while lies at the upper end of the predictions from the non-ideal MHD model.

\begin{figure*}[h]
\centering
\includegraphics[width=\textwidth]{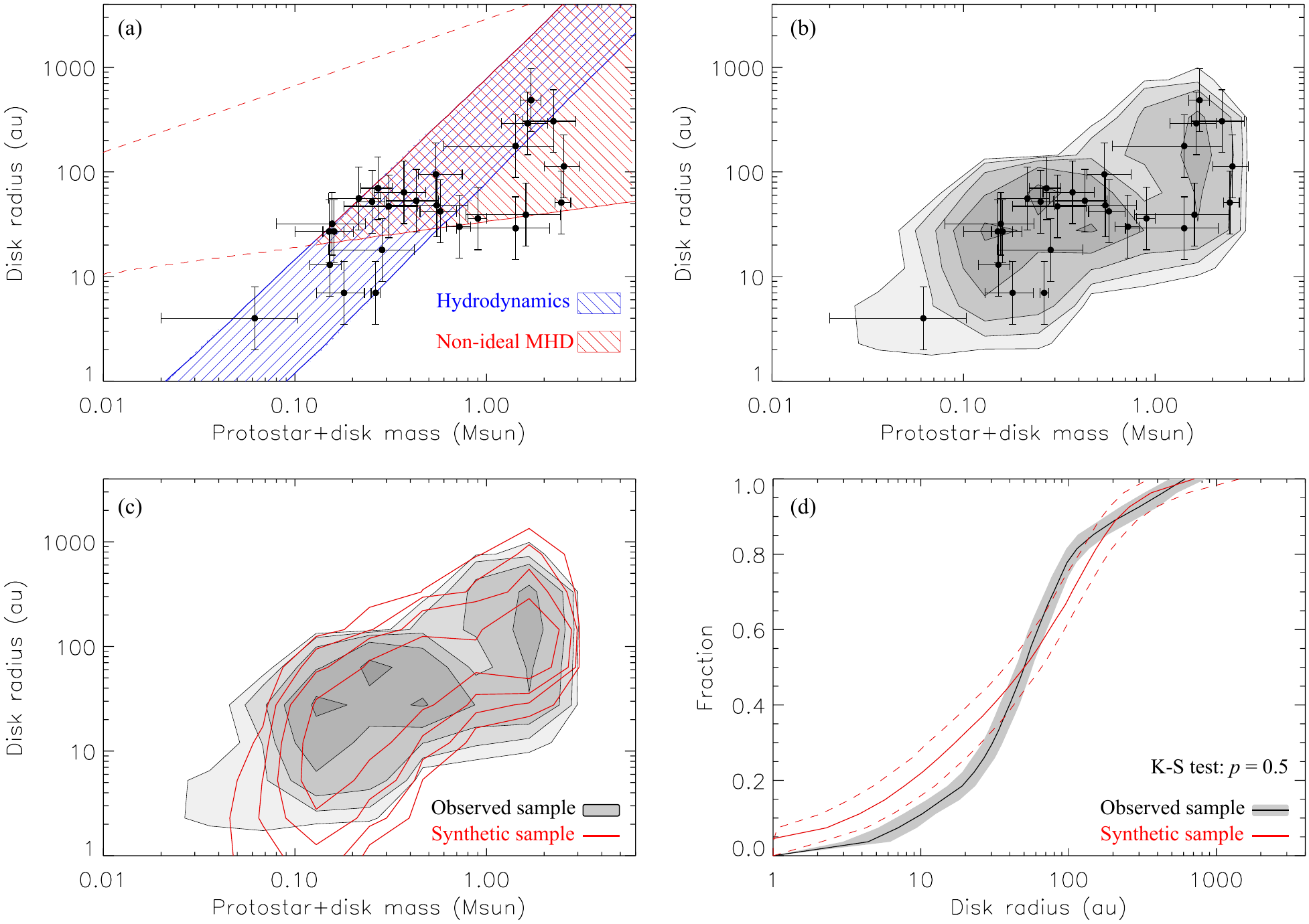}
\caption{(a) Observed disk radius and protostar+disk mass distribution in our sample (data points; the same as Fig.~\ref{fig:ob_rd}) in comparison with the theoretical predictions. Blue and red shaded areas present the theoretically predicted relationship between disk radius and protostar+disk mass from the hydrodynamic (Eq.~\ref{eq:Rhydro}) and non-ideal MHD models (Eq.~\ref{eq:incli} and \ref{eq:MHD}), respectively. The non-ideal MHD solutions in principle extend toward the left-hand side, as delineated with red dashed lines. However, those solutions implying more angular momentum than that available in initial dense cores, so they are physically irrelevant. (b) Probability distribution of disk radii and protostar+disk masses of our sample sources (gray shaded area and contours) overlaid with the observed data points. From outer to inner contours, they enclose 99\%, 92\%, 87\%, 60\%, and 27\% of the total number of the sample sources, respectively. (c) is the same as (b) but for the comparison of the probability distributions between our observed and synthetic (red contours) samples. The synthetic sample is generated with the observed protostar+disk mass distribution and theoretical disk radii from hydrodynamics and non-ideal MHD. (d) Cumulative distributions of the disk radii of our observed (black) and synthetic (red) samples. Gray shade area and red dashed lines delineate the corresponding uncertainties. Similar to Fig.~\ref{fig:Rd}, the cumulative distributions and their uncertainties are inferred from 1000 iterations of varying data points within their uncertainties. The K-S test on the two distributions results in a $p$ value of 0.5.}
\label{fig:rd}
\end{figure*}

Using Eq.~\ref{eq:Rhydro}--\ref{eq:MHD}, the theoretical disk radii from the hydrodynamic and non-ideal MHD models are computed and plotted in Fig.~\ref{fig:rd}a.
For a given protostar+disk mass, the range of the theoretical disk radii from the hydrodynamic model is determined by the range of $\beta$, which is based on the observationally inferred angular momentum profile in dense cores (Fig.~\ref{fig:jr}), and those from the non-ideal MHD model is determined by the magnetic field orientations.
The theoretical disk radius from the non-ideal MHD model reaches the minimum and maximum when the external magnetic field is vertical or horizontal with respect to the disk plane, respectively.

We note that with Eq.~\ref{eq:MHD}, the non-ideal MHD predictions in principle extend toward smaller protostar+disk masses, and have theoretical disk radii larger than those from hydrodynamics.
These solutions require more angular momentum than that available in initial dense cores to reach the equilibrium state. 
Therefore, they are physically irrelevant \citep{LeeYN24} and are not considered here.

Figure \ref{fig:rd}a shows that the theoretical predictions from the hydrodynamic and non-ideal MHD models presented in Sec.~\ref{sec:theory} can explain the distribution of the observed disk radii and protostar+disk masses.
The disk radii with protostar+disk masses lower than 0.3 $M_\sun$ can generally be explained with the hydrodynamic model. 
For the disks with protostar+disk masses higher than 0.3 $M_\sun$, their radii gradually deviate from the prediction from the hydrodynamic model, 
and becomes too small to be explained by the hydrodynamic model when the  protostar+disk masses exceed 0.6 $M_\sun$.
The disk radii of these higher-mass sources can be explained with the non-ideal MHD model.

For a quantitative comparison between the theoretical predictions and observations, 
we generated synthetic samples using Eq.~\ref{eq:Rhydro}--\ref{eq:MHD} and the observed protostar+disk masses. 
For each iteration, 
we created 27 synthetic sources, the same number as our observed sample, 
and their protostar+disk masses were adopted from the observational measurements and randomly varied within the observational uncertainties. 
Given their protostar+disk masses, values of $\beta$, also randomly varied within the uncertainty, were assigned to them using Eq.~\ref{eq:beta}.
The observations of the magnetic fields on 1000 au and 0.1 pc scales in protostellar sources suggest that the magnetic field orientations on both scales are approximately random \citep{Hull14,Yen21a}.
Thus, a random orientation of the magnetic field with respect to the disk plane was also assigned to each synthetic source. 
Finally, the theoretical disk radii from the hydrodynamic and non-ideal MHD models were computed for the synthetic sources with Eq.~\ref{eq:Rhydro}--\ref{eq:MHD}. 
If the theoretical disk radius from the non-ideal MHD model was smaller than that from the hydrodynamic model, 
it was assigned to that synthetic source, meaning that its disk radius is regulated by magnetic braking and non-ideal MHD effects. 
If the theoretical disk radius from the hydrodynamic model was smaller, 
it was assigned to that synthetic source, meaning that magnetic braking is not efficient and hydrodynamics is more important in that source (Sec.~\ref{sec:theory}).
We repeated this process 1000 times. 

We mapped the observed data points and synthetic sample to two-dimensional probability distributions (e.g., Fig.~\ref{fig:rd}b).
Figure \ref{fig:rd}c compares the resultant probability distributions from the observations and the synthetic sample.
We note that our sampling of protostar+disk mass is not uniform, as two local maximums are seen in the observed probability distribution.
Nevertheless, our synthetic sample is generated based on the observed protostar+disk masses in our sample, and thus a fair comparison can be made.
Figure \ref{fig:rd}c shows that the theoretical prediction approximately matches the observed probability distribution.

Figure \ref{fig:rd}d compares the cumulative distributions of the disk radii in our observed and synthetic sample. 
The two dashed lines enclose the 68\% probability of the distribution based on our 1000 iterations. 
The K-S test on the two observed and synthetic distributions results in a $p$ value of 0.5. 
This suggests that the two distributions are statistically indistinguishable, 
and the combination of the theoretical predictions from the hydrodynamic and non-ideal MHD models can indeed explain the observed disk radius distribution. 
This theoretical model also predicts the rarity of large ($>$100 au) disks and the presence of several very small ($<$10 au) disks, 
and both are less than 20\% in the synthetic sample.

We note that our theoretical model may underestimate disk radii for low mass protostars with protostar+disk masses of $\lesssim$0.1--0.2 $M_\sun$. 
Although statistically it is not significant, 
our synthetic sample has a higher fraction of disks with radii smaller than 10--20 au (Fig.~\ref{fig:rd}c and d). 
This is likely due to our assumption of a SIS and the adopted radial profile of $\beta$.
The density increases to infinite at the center in a SIS. 
It is inconsistent with observed dense cores typically having a flatter density profile at small radii initially \citep{Alves01}. 
Besides, in a SIS, the initial radius enclosing 0.1 $M_\sun$ is smaller than 0.01 pc, 
while the rotation in dense cores at radii smaller than 0.01 pc is not well explored observationally (Fig.~\ref{fig:jr}). 
Therefore, it is possible that $\beta$ is underestimated with Eq.~\ref{eq:beta} in the low-mass regime, 
leading to potential underestimate of disk radii for low-mass protostars in our synthetic sample.
This also causes the dependence of the disk radii on the protostar+disk masses with a power-law index of 1.2$\pm$0.2 in our synthetic sample to be steeper than the observations.

\section{Discussion}\label{sec:discussion}
Our results suggest that the combination of theoretical predictions from the hydrodynamic and non-ideal MHD models in \citet{LeeYN21,LeeYN24} can explain the observed disk radius distribution (Fig.~\ref{fig:rd}). 
In this theoretical model, 
the disk growth is controlled by hydrodynamics in the low protostar+disk mass regime ($\lesssim$0.3 $M_\sun$), such as protostellar sources at the early stage of gravitational collapse of a dense core or in the formation process of low-mass stars. 
The exact mass threshold depends on the rotational energy in individual dense cores. 

Magnetic braking is due to the magnetic tension. 
In this low-mass regime (or early evolutionary stage), the magnetic fields have not been severely twisted because the gas rotation remains slow, 
as suggested by the radial profile of rotational to gravitational energy (Eg.~\ref{eq:beta}).
Thus, magnetic braking is not dynamically important, with its time scale longer than that of advection of angular momentum from the infalling protostellar envelope \citep{LeeYN21}.\footnote{The advection time scale ($\tau_{\rm adv}$) is proportional to the infall time as $\tau_{\rm adv} = \frac{r}{u_r}$, where $r$ is the radius and $u_r$ is the infalling velocity. The magnetic braking time scale ($\tau_{\rm br}$) is inversely proportional to magnetic field strength as $\tau_{\rm br} = \frac{\rho u_\phi h}{B_z B_\phi}$, where $\rho$ is the density, $u_\phi$ is the azimuthal velocity, $h$ is the scale height, and $B_z$ and $B_\phi$ are vertical and azimuthal magnetic field strengths, on the disk scale \citep{LeeYN21}.} 
As a result, the disk are small ($<$10 au) due to the low rotational energy of the collapsing material and can grow rapidly with the proceeding collapse. 

As the magnetic fields being dragged inward and twisted with the proceeding collapse and increasing rotation,
azimuthal magnetic fields are quickly developed until the balance between induction and diffusion is reached.
This makes magnetic braking more efficient.
Consequently, the disk growth is suppressed with its radius determined by the balance between the angular momentum carried inward by mass accretion and removed by magnetic braking.
The central mass--disk radius relation from the non-ideal MHD model describes this equilibrium state (Eq.~\ref{eq:MHD}), where the time scales of advection of angular momentum and magnetic braking are comparable \citep{LeeYN21}.

In Fig.~\ref{fig:rd}a, those protostellar sources in the higher-mass regime ($\gtrsim$0.3 $M_\sun$) are likely in this non-ideal MHD dominant regime. 
In the lower-mass sources located below this central mass--disk radius relation from the non-ideal MHD model, 
such the equilibrium is not reached, and their advection time scale is shorter than magnetic braking time scale.
If their protostar+disk systems grow and develop substantial azimuthal magnetic fields with increasing rotation, 
they may approach this equilibrium state and evolve to the non-ideal MHD dominant regime. 
For protostars with low final masses, 
they may stay in the hydrodynamics dominant regime throughout the star formation process, 
and their disk growth can be fully controlled by hydrodynamics.

In the non-ideal MHD dominant regime, most of our sample disks are found near the lower end of the model predictions (the red shaded area in Fig.~\ref{fig:rd}a).
This is because non-ideal MHD regulated disks become several hundred au in radius only if the magnetic fields are largely misaligned to the disk rotational axes. 
Given random orientations of magnetic fields, 
statistically the number of disks with the magnetic fields largely misaligned ($>$80$\arcdeg$) to the rotational axes is limited \citep[$<$20\%;][]{Yen21a}. 

Although our model can explain the observed disk radius distribution (Fig.~\ref{fig:rd}), the expected change in the slope of the protostellar mass--disk radius relation from the hydrodynamics to non-ideal MHD dominant regimes is not yet observed in our current data with the limited sample size, as the single power-law function can describe the data better than the double power-law function (Fig.~\ref{fig:ob_rd}). 
Future observations to better sample the parameter space are needed to further examine this scenario.
In addition, we note that detecting Keplerian rotation in small ($<$10--20 au) disks to measure their central protostellar masses remains observationally challenging.  
Observations targeting small disks and measuring their central protostellar masses are critical to test this model and examine if any sources, such as higher-mass protostars with small disks, are located outside the parameter space predicted from this model.

\section{Summary}
To investigate the mechanisms determining disk formation and growth,
in this paper, we compare the observed disk radii with the theoretical predictions from the hydrodynamic and non-ideal MHD models in \citet{LeeYN21,LeeYN24}.
We compile a list of observationally inferred specific angular momentum profiles in dense cores and a sample of 27 Class 0 and I single protostars with resolved continuum disks and dynamically determined protostellar masses from the literature.

Using these observed protostellar masses and the rotational profile in dense cores, 
we compute theoretical disk radii from the hydrodynamic and non-ideal MHD models, and generate synthetic samples for a statistical comparison with the observed disk radius distribution. 
The magnetic field orientations are assumed to be random in the synthetic samples.
Our theoretical model can explain the observed disk radius distribution. 

In this model, the disk growth is controlled by hydrodynamics in the low-mass regime, approximately when the central protostar+disk mass is below 0.3 $M_\sun$. 
With the proceeding collapse, the central protostar+disk system grows and develops azimuthal magnetic fields.
Magnetic braking becomes dynamically important and starts to suppress the disk growth, when the system exceeds the mass threshold, which depends on the rotational energy in the initial dense core.
Then the disk growth is determined by non-ideal MHD, which regulates the efficiency of magnetic braking. 
This theoretical model with hydrodynamics and non-ideal MHD dominant in the low- and high-mass regimes naturally explain the rarity of large ($>$100 au) protostellar disks and the presence of very small ($<$10 au) protostellar disks. 
It also predicts that the majority of protostellar disks have radii of a few tens of au, as observed. 
Future observations with larger samples to measure the slope of the protostellar mass--disk radius relation in the low and higher mass regimes are essential to further examine this model.

\begin{acknowledgements}
We would like to acknowledge the entire ALMA eDisk team for enabling the detection of a substantial sample of protostellar disks, which has significantly contributed to our statistical study.
H.-W.Y.\ acknowledges support from the National Science and Technology Council (NSTC) in Taiwan through grant NSTC 113-2112-M-001-037-, 113-2112-M-001-035-, and 112-2124-M-001-014- and from the Academia Sinica Career Development Award (AS-CDA-111-M03).
Y.-N.L.\ acknowledges funding from NTSC in Taiwan (NSTC 112-2636-M-003-001) and the grant for Yushan Young Scholar from the Ministry of Education in Taiwan.
\end{acknowledgements}


\end{document}